# 155-day Periodicity in Solar Cycles 3 and 4

November 3, 2009

J. M. Vaquero (1,2), R. M. Trigo (2,3), M. Vázquez (4,5) and M. C. Gallego (6)

(1) Departamento de Física Aplicada, Escuela Politécnica, Universidad de Extremadura, Cáceres, Spain.
(2) CGUL-IDL, Universidade de Lisboa, Lisbon, Portugal.
(3) Departamento de Eng. Civil da Universidade Lusófona, Lisbon, Portugal.
(4) Instituto de Astrofísica de Canarias, La Laguna, 38200 Spain.
(5) Departamento de Astrofísica, Universidad de La Laguna, 38205 La Laguna, Spain
(6) Departamento de Física, Universidad de Extremadura, Badajoz, Spain.

Abstract
The near 155 days solar periodicity, so called Rieger periodicity, was first detected in solar flares data and later confirmed with other important solar indices. Unfortunately, a comprehensive analysis on the occurrence of this periodicity during previous centuries can be further complicated due to the poor quality of the sunspot number time-series. We try to detect the Rieger periodicity during the solar cycles 3 and 4 using information on aurorae observed at mid and low latitudes. We use two recently discovered aurora datasets, observed in the last quarter of the 18th century from UK and Spain. Besides simple histograms of time between consecutive events we analyse monthly series of number of aurorae observed using different spectral analysis (MTM and Wavelets). The histograms show the probable presence of Rieger periodicity during cycles 3 and 4. However different spectral analysis applied has only confirmed undoubtedly this hypothesis for solar cycle 3.

PACS: 96.50.Wx, 96.60.-j, 96.60.Q-, 92.60.hw

# 1 Introduction

Rieger et al. (1984) and Dennis (1985) detected a periodicity of approximately 158 days in γ and X rays flare data taken by the SMM and GOES satellites during the solar cycle 21. Later, this solar periodicity has also been detected in different epochs using several solar activity indices: Sunspot areas (Carbonell and Ballester 1990, Lean 1990), 10.7 cm radio flux (Lean and Brueckner 1989), flares (Ichimoto et al. 1985, Chowdhury et al. 2008, Dimitropoulou et al. 2008), Mount Wilson Sunspot Index (Ballester et al. 2004), Interplanetary Magnetic Field strength (Cane et al. 1998), neutrino flux (Sturrock et al. 1997) but not in plage index (Lean and Brueckner 1989). This periodicity has been better detected during solar maximum (Ballester et al. 2004) and is also stronger in strong solar activity cycles. However, this could simply reflect the lack of sunspots, or related phenomena, for other epochs of the solar 11-year cycle.

The periodicity has also an asymmetrical character showing different power in the two hemispheres (Knaack et al. 2005), behaviour also found in the solar activity itself (Carbonell and Ballester 1990).

The majority of published results have been obtained using datasets restricted to the last 10 cycles due to the shortage of representative solar indices for previous times. Silverman (1990) and Ballester et al. (1999) demonstrated that Rieger periodicity was present in the Sun in historical epochs. While Silverman (1990) managed to show different historical auroral series with the above mentioned periodicity, Ballester et al. (1999) used a new record of the solar activity: the number of sunspot groups. This index has been reconstructed by Hoyt and Schatten (1998) from 1610 to 1995. A wavelet analysis of this time series shows clearly that an episode of a periodicity close to 158 days occurred during the 18th century corresponding to the solar cycle number 2. Unfortunately, the analysis is not valid for the entire 18th century since the above mentioned index is of relatively poor quality for most of the 18th century and, especially during the last quarter, a period that corresponds to the solar cycles number 3 and 4 (Vaquero 2007). As a result on these caveats, the Rieger periodicity has not been detected for most of the 18th and 19th centuries, being only confirmed for 5 (out of 13) solar cycles between 1730 and 1890. Moreover, Usoskin et al. (2001) suggested from the few information available that solar cycle 4 was in fact a superposition of two cycles: a normal 10-year long cycle between 1784 and 1793 followed by a short and weak cycle in 1793-1800. Arguments pro and against the "lost cycle" theory have been proposed (Usoskin et al. 2002, Krivova et al. 2002, Usoskin et al. 2003). Zolotova and Ponyavin (2007) have suggested recently that the unusual length of the solar cycle 4 could be explained by outstanding phase asynchrony between Northern and Southern hemispheric activities. Recently, Usoskin et al. (2009) have shown clearly, based on the systematic occurrence of sunspots at high solar latitudes in 1793–1796, that a new cycle has started indeed in 1793 and that such short-cycle has been neglected in the traditional Wolf sunspot series.

Wavelet analyses performed by Krivova and Solanki (2002) shows that the Rieger 155-d has a common origin with the 1.3 year periodicity, present in helioseismic data (Howe et al. 2000). Such 1.3 period was also detected by Silverman and Shapiro (1983) using auroral data from 1721 to 1943. However, this periodicity was not verified by Basu and Antia (2001) and recently Howe et al. (2007) failed to detect it after 2001. Knaack et al. (2005) provide evidence for a multitude of quasi-periodic oscillations of the photospheric magnetic field in this temporal range, suggesting that the 1.3 year period could be related to large-scale magnetic surges appearing in the 30-55 latitude belts during the maxima of the solar cycles 21-23.

Krivova and Solanki (2002) suggest that both 1.3 year and 155 days could be harmonics of the 11year cycle, without excluding a link with the solar dynamo via changes in the solar rotation. Analyzing solar flare data Temmer et al. (2005) proposes that the appearance of active periods around 155 days might result from the superposition of predominant periods of 24 and 28 days, related with the solar rotation period.

Recently, two new sets of historical aurorae data, recorded at mid latitudes and covering the late 18th and early 19th centuries have been published (Vaquero, Gallego and García 2003, Harrison 2005). The aim of this work is to evaluate if the Rieger periodicity is also present in these datasets during the maxima of the solar cycles 3 and 4 (1777–1789), an epoch previous to the onset of the Dalton Minimum of solar activity.

In order to place our working window in a temporal perspective, we plot in Fig. 1 the group sunspot number for the period 1730–2000, indicating in the upper bars those epochs where the Rieger periodicity has been clearly detected by other authors using a number of different methodologies and data-sets: (1) Sunspot Area (Lean 1990), (2) Group Sunspot Number (Ballester et al. 1999), and (3) Aurorae (Silverman 1990). Note that the Sunspot Area data-set starts only in 1874. There is a data-set with previous data (from 1832 until 1868) derived by Vaquero et al. (2002), however only longer (about 330 days) and shorter (30-50 days) cycles have been recovered, lacking the 150-160 days periodicity (Vaquero, Gallego and Sánchez-Bajo 2003). We have marked also the presence of the Rieger periodicity in solar cycles number 3 and 4, that will be discussed further in the following sections.

## 2 Data

The 19th-Century Spanish physicist Rico Sinobas compiled a catalogue of aurorae observed from the Iberian Peninsula during the period 1700–1855 (40 auroral events). The data recorded by Rico Sinobas is consistent with respect to expected characteristics of colour, azimuth, and distribution with respect to the phase of the moon (Vaquero, Gallego and García 2003). Similarly to other historical aurorae datasets, Rico Sinobas dataset shows a notable gap in the time series of observed aurorae during the Dalton minimum.

The 1771–1813 diary of Thomas Hughes from Stroud, Gloucestershire (51.75ºN, 2.22º W), has provided a rich source of historical meteorological data. Additionally, it describes 71 nights on which the aurora was seen, between 19 February 1771 and 13 October 1805 inclusive. About 90% of Hughes' aurora observations occurred between 1771 and 1789. These observations provide information on the incidence of the aurora in southern UK, before the widespread nuisance of light pollution. Harrison (2005) published the auroral observations recorded in the dairy of Hughes recently.

We must note that there are two different interplanetary origins of geomagnetic storms (and terrestrial aurorae) that vary strongly with the phase of the 11-year solar cycle. The interplanetary manifestations of coronal mass ejections (CMEs) cause intense geomagnetic storms around solar maximum (Gonzalez et al. 2002). However, high speed streams of solar wind from coronal holes control the interplanetary medium activity during solar minimum (Gonzalez et al. 1999). According to the distribution of auroral occurrence compiled by Rico Sinobas and observed by Hughes (see Fig. 2), we can infer that these aurorae are associated to the first type of interplanetary origin.

Available information on the number of sunspot groups and the aurorae recorded in the catalogues of Rico Sinobas and Hughes during the period 1777–1789 can be observed in Fig. 2 (that includes the consecutive maxima of the solar cycles 3 and 4). Due to poor quality of the sunspot group data, Rieger periodicity was not detected previously in this series in the analysis made by Ballester et al. (1999). Nevertheless, a close visual inspection of the dates of consecutive auroral events reveals several temporal intervals consistent with the Rieger periodicity (depicted in the upper part of Fig. 2 with the small arrows).

As the existence of these time intervals between auroral events is evident from Fig. 2, we would like to verify the statistical significance of the frequency with regard to the different intervals between consecutive events. With this aim, we have constructed a histogram for both sets of observations, grouping the intervals between events in bins with 20-day width (Fig. 3). We should expect significant peaks in the 10-day bin and in the 30-day bin because intense geomagnetic storms (that tend to produce auroral displays seen on successive nights) and because of the 27-day recurrence tendency of some geomagnetic storms respectively.

In both sets of observations (Fig. 3a and 3b), one can appreciate a peak in the frequency of events with intervals near to Rieger periodicity (bin centred in 150 days). Nevertheless, the peak relative to the Rico Sinobas catalogue is more prominent. We have estimated the statistical significance of our results with a Monte-Carlo simulation that have generated 200 000 random time series constrained with a similar amount of auroral events. We show in Fig. 3, for every bin category of the corresponding histograms (obtained for each random sample), the mean value (thick line) and the corresponding values associated to ±1σ (thin line) and ±2σ (dashed line) intervals. The histogram constructed using Sinobas' data shows that the bin centered in 150 days is the only that surpass the +2σ limit. Nevertheless, in the case of Hughes data, three bins (10, 30, and 150 days) are surpassing the +2σ limit. Therefore, a first inspection of these auroral data shows the probable presence of Rieger periodicity during these epochs.

It should be stressed that some dates of auroral display appear in both catalogues compiled by Hughes and Rico Sinobas (e.g. 1780-1781). However, during some years the approximately 155-day intervals between auroral events in the two catalogues appear to be almost "out of phase" (e.g. 1788-1789). Although the sites in the Iberian Peninsula are located at lower magnetic latitudes than Stroud, Gloucestershire, geomagnetic storms that were sufficiently strong to produce auroral displays seen in Spain or Portugal are capable of producing an auroral display that would also have been seen in Central England under favourable meteorological conditions (i.e. as the auroral oval moves southwards).

Therefore, some of the discrepancies between the two catalogues mentioned above may result from unfavourable meteorological conditions in England. In order to evaluate the level of dataset dependency of our results we have repeated the analysis but now applied to the combined dataset that resulted from the aggregation of Sinobas and Hughes individual datasets (Figure 4). Results are not so clear cut as those obtained for the individual datasets. The combined dataset reveals that the bin centered in 150 days surpasses the +2σ limit for the data of solar cycle 3 (Fig. 4a). However, this bin only surpasses the +1σ limit for solar cycle 4 (Fig. 4b) and also for the complete period of the analysis (Fig. 4c).

# 3 Spectral analysis

The accomplishment of a spectral analysis in the available datasets presents important technical problems. If we take a daily timescale to construct our time series (as appears in Fig. 2), the majority of the daily values of observed aurora will be zero. In fact, the same problem arises when one attempts to perform a spectral analysis of the sunspot number during Maunder Minimum (Frick et al. 1997). To overcome this difficulty when performing spectral analysis, we have constructed monthly series of number of observed aurorae. However, we must acknowledge that this approach implies the loss in spectral resolution and difficulting the separation of the 155 days peak from the semiannual cycle of 180 days.

Fig. 5 shows the MTM (Multi-Taper Method) spectra of the monthly time series of both auroral datasets (upper panel) and of sunspot numbers (lower panels). We have used the International sunspot number, $R_Z$, (middle panel) and the Group Sunspot Number, $R_G$, (lower panel) as indices for solar activity. For the period 1777–1781 (left side, Fig. 5), the MTM spectrum of the monthly number of aurora recorded in the catalogue of Rico Sinobas shows a statistically significant (at 1% significance level) cycle of 145 days. This cycle also is observed in the MTM spectra of $R_Z$ and $R_G$ for the period 1771–1781. For the period 1779–1789 (right side, Fig. 5), the MTM spectrum of the monthly number of aurora recorded in Hughes's observations shows a statistically significant (at 1% significance level) cycle of 180 days corresponding to the semiannual geomagnetic cycle and the first harmonic of 90 days. The MTM spectra of $R_Z$ and $R_G$ for the period 1779–1789 do not show any significant cycle.

In order to further confirm the existence of the Rieger periodicity in these datasets we have also applied a wavelet power spectrum analysis (Torrence and Compo 1998). This technique is particularly useful for non-stationary time series that are dominated by periodicities that change in time. Results obtained with the Hughes dataset can be appreciated in Fig. 6. The monthly number of auroral observations recorded by Hughes between 1770 and 1806 can be seen in Fig. 6a, while the corresponding wavelet power spectrum of the data can be appreciated in Fig. 6b. It is particularly worth mentioning the presence of a broad 150-180 day and 1.4 year periodicities, although restricted to the first half of the available time series. Data from early 19th century reveals very few auroral events (Fig. 6a) therefore hampering the appearance of any statistically significant periodicities in the wavelet power spectrum for that period (Fig. 6b). This mixed behaviour cannot be observed with a classical power spectrum analysis (e.g. Fast Fourier Transform), and that caveat helps to explain the barely significant peaks associated with the 150-180 day and 1.4 year periodicities in the global wavelet power spectrum (black line) shown in Fig. 6c.

## 4 Conclusions

We have detected the Rieger periodicity in one auroral dataset of the last quarter of the 18th century (1777-1781) both in the histograms of time intervals between consecutive auroral events as well as in the spectrum of the monthly series of the number of auroral events. However we acknowledge that we could not confirm so clearly the same periodicity between 1779 and 1789.

The above mentioned intermittency and the drift of the main observing periodicity is one of the characteristics of the process under study. We must stress that the solar origin of aurorae compiled by Rico and observed by Hughes is not associated to coronal holes but mostly to CMEs (and, probably, solar flares). Therefore, both auroral data-sets are particularly appealing to search for periodicities related to great solar actives regions, whereas most auroral datasets combine aurorae caused by other mechanisms. Naturally it would be relevant to locate the solar physical mechanism responsible for the appearance of the 155-day periodicity. In particular it is crucial to understand if these phenomena take place deep under in the convection zone or just close to the final zone of emergence of active regions (Bai and Sturrock 1987, Dimitropoulou et al. 2008). However, prior to locate and understand the physical mechanism associated with this Rieger frequency one must guarantee that this is not merely a statistic artifact.

Rieger periodicity has been detected in the more recent solar cycles, but also, through indirect forms in some ancient cycles. It is important to verify the consistency of this periodicity during the last 250 years. In this context, the permanent search of ancient solar activity proxy information may offer relevant information about the reproducibility of Rieger cycle in all solar cycles.

We acknowledge that time series on the number of aurora observed from a single location should be used carefully as a proxy of solar activity in historical epochs due to the influence of local meteorology (visibility conditions). Nevertheless, we believe that this work highlights the potential of using ancient auroral records for the study of the characteristics of the solar activity during epochs in which the group sunspot number presents a poor coverage (Vaquero 2007, Vaquero and Vázquez 2009).

**Acknowledgements.** This work was partially supported by the Ministerio de Ciencia e Innovación of the Spanish Government (AYA2008-04864/AYA). The authors are in debt to David M. Willis who provided important comments that improved our initial manuscript.

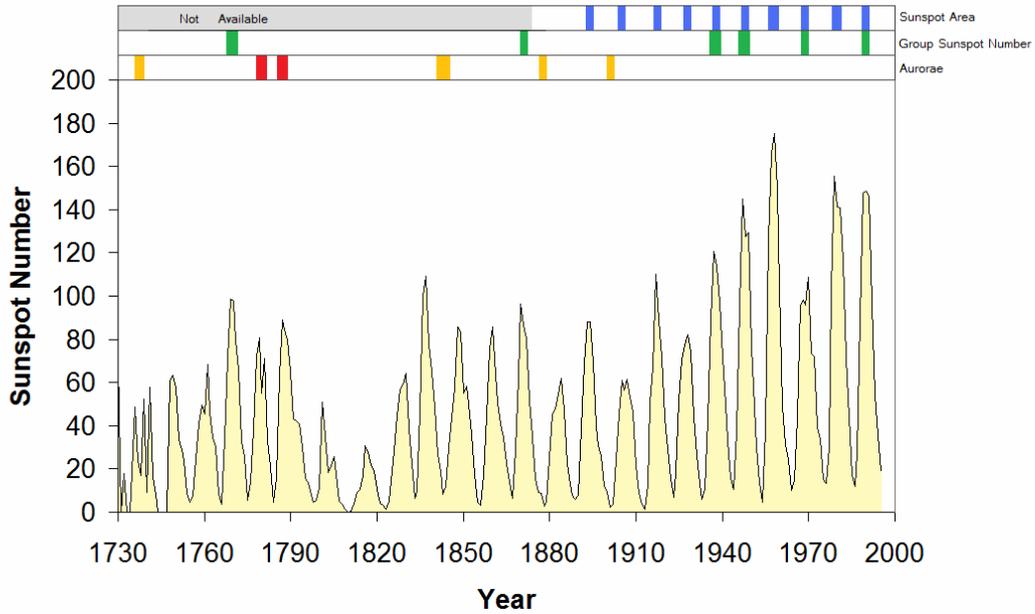

Figure 1: Group Sunspot Number for the period 1730–2000 and epochs where the Rieger periodicity has been clearly detected using (blue) Sunspot Area (Lean, 1990), (green) Group Sunspot Number (Ballester et al., 1999) and (yellow) Aurorae (Silverman, 1990). The two cycles (3 and 4) dealt in this work with the new Aurorae catalogues are highlighted in red.

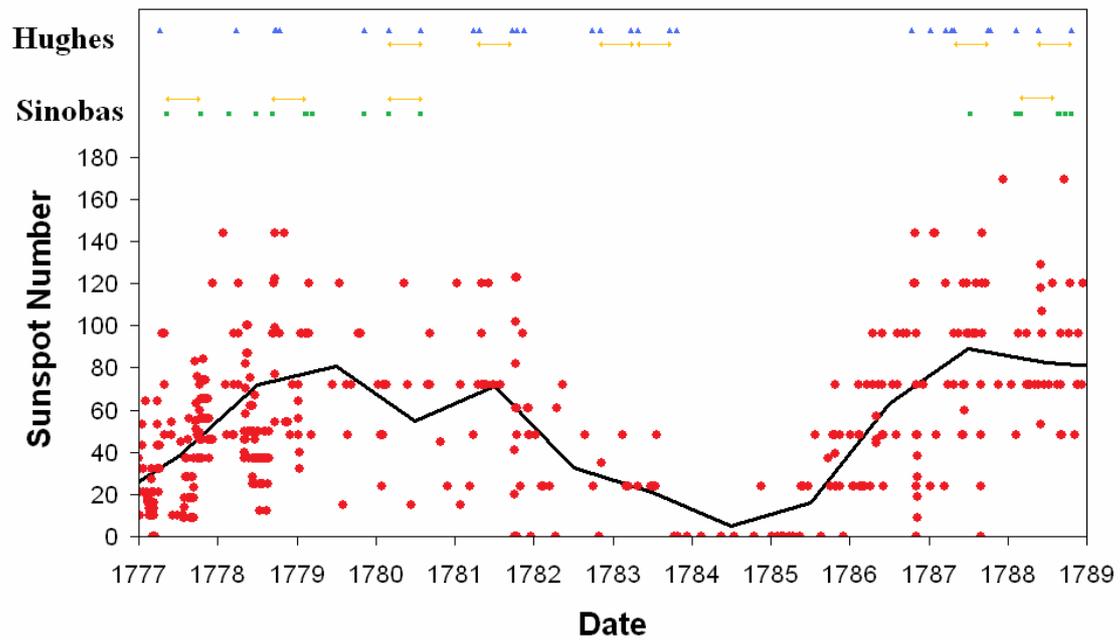

Figure 2: Daily (dots) and monthly (line) sunspot numbers during the period 1777-1789 and auroral events recorded by Rico Sinobas and Hughes. The small arrows depict consecutive auroral events separated by Rieger period (155 days approximately).

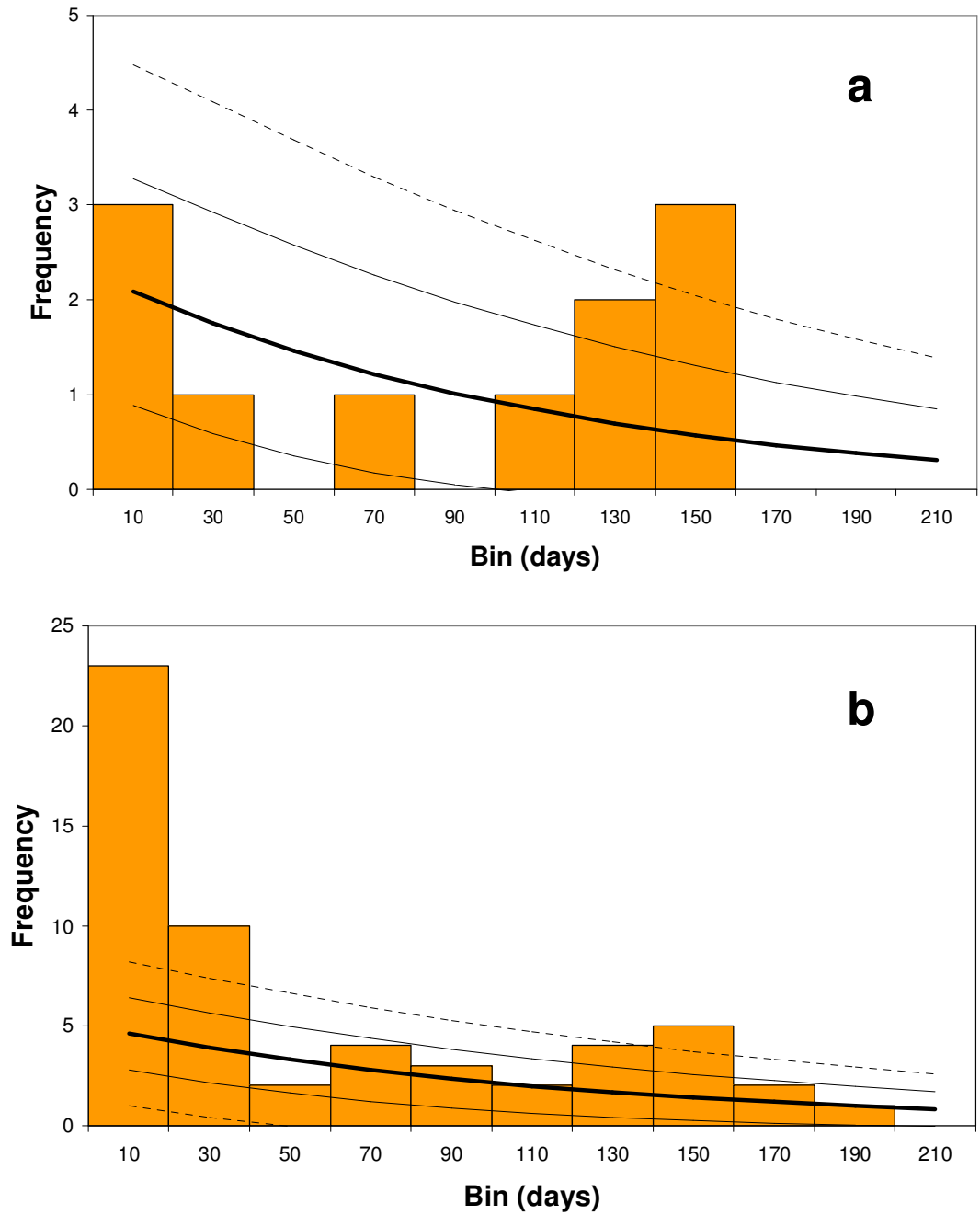

Figure 3: Frequency histograms on the number of cases as a function of the lapse time between auroral events for the (a) Rico Sinobas and (b) Hughes catalogues. Also shown are the mean value (thick line) and the corresponding values associated to ±1σ (thin line) and ±2σ (dashed line) intervals obtained with a Monte-Carlo simulation that have generated 200.000 random time series (constrained with a similar amount of auroral events).

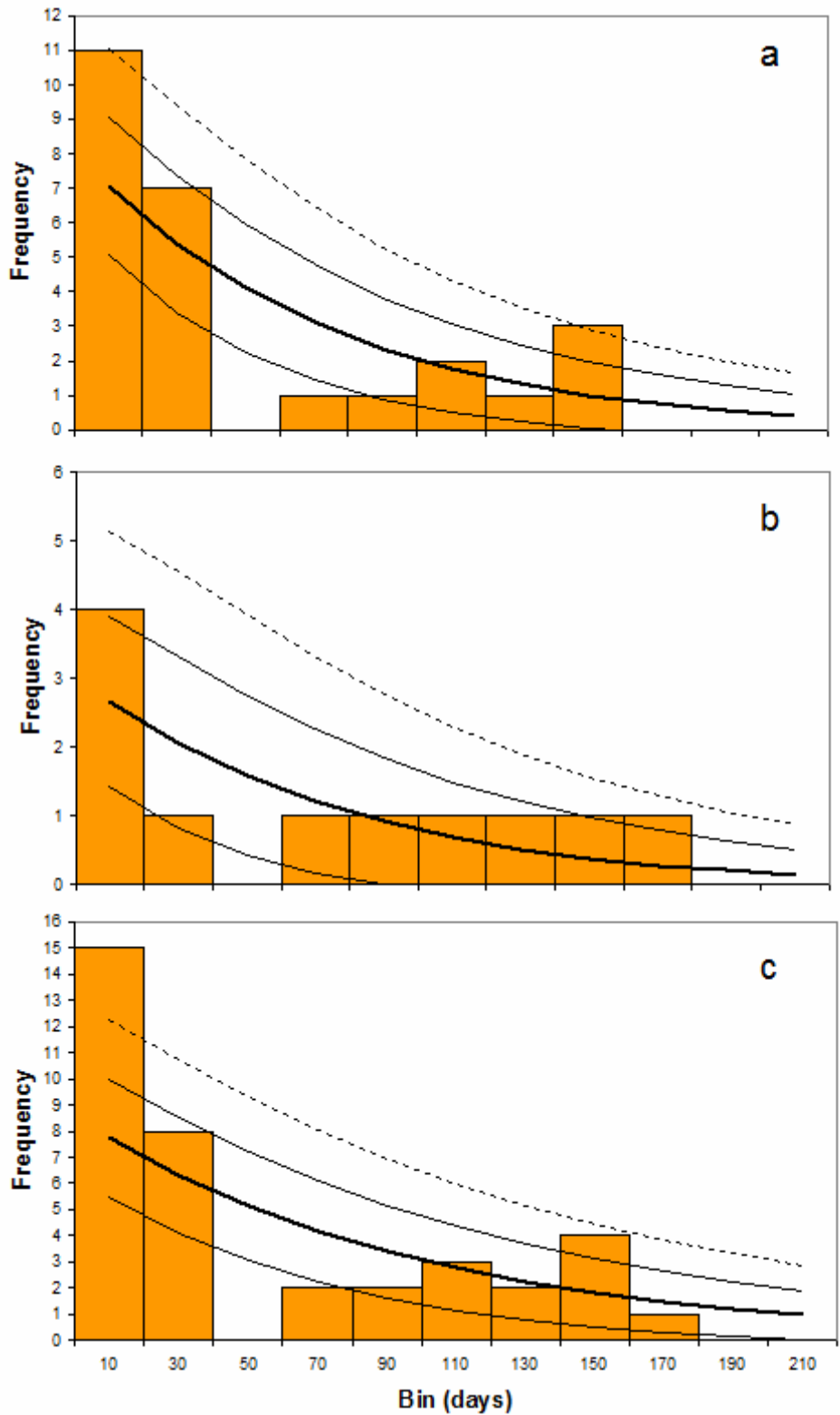

Figure 4: Frequency histograms on the number of cases (using the combined dataset from Rico Sinobas and Hughes catalogues) as a function of the lapse time between auroral events for (a) solar cycle 3, (b) solar cycle 4, and (c) solar cycles 3 and 4. Significance levels as in Figure 3.

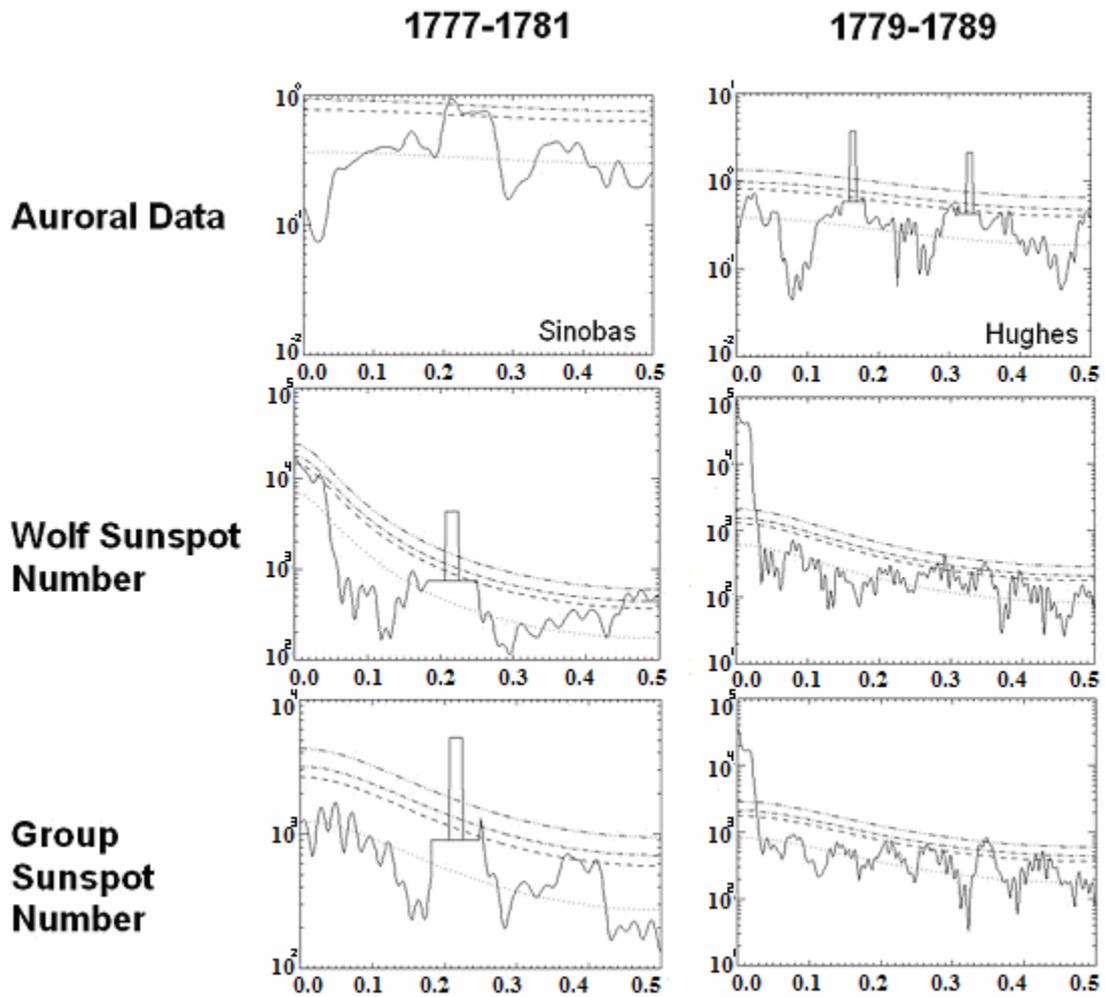

Figure 5: MTM spectral analysis of monthly number of aurorae used in this work (upper), Wolf sunspot number (middle) and Group sunspot number (lower) during the periods 1777–1781 (left) and 1779–1789 (right).

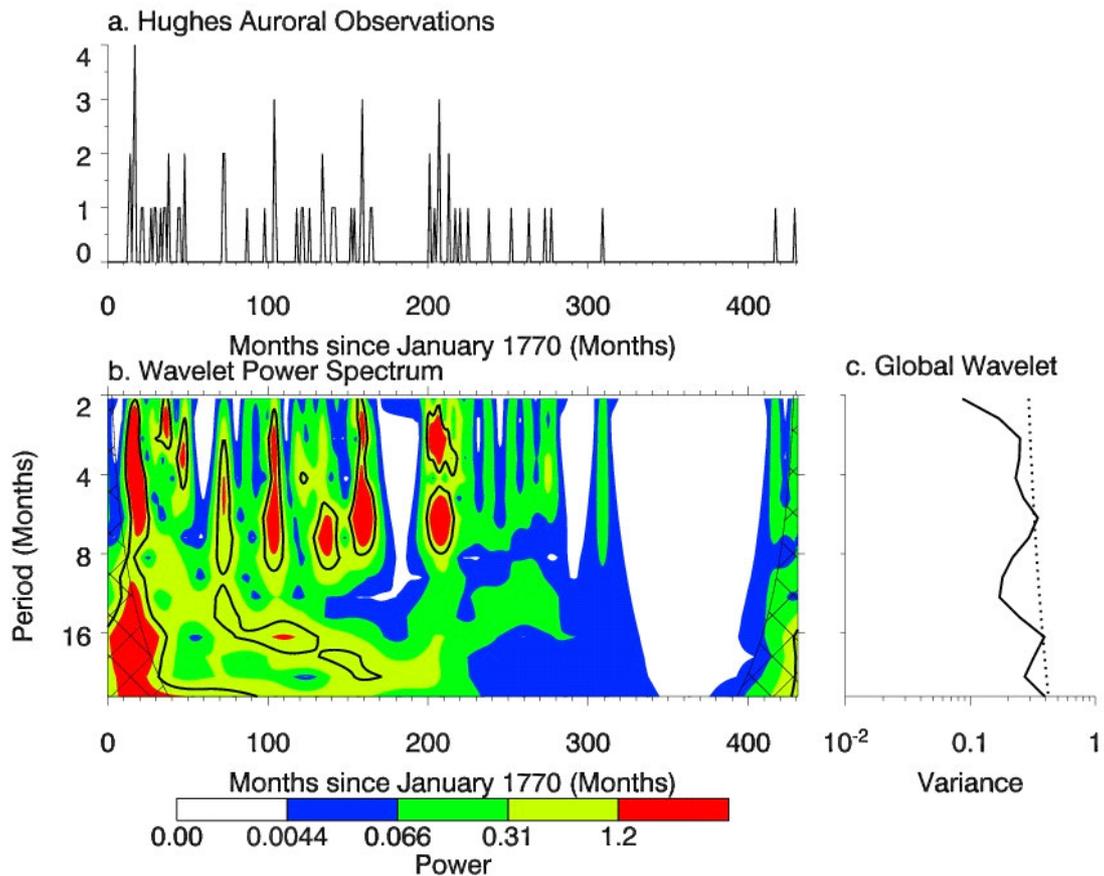

Figure 6: (a) Hughes auroral observations between 1770 and 1806. (b) The wavelet power spectrum. The contour levels are chosen so that 75%, 50%, 25%, and 5% of the wavelet power is above each level, respectively. The cross-hatched region is the cone of influence, where zero padding has reduced the variance. Black contour is the 5% significance level, using a red-noise background spectrum. (c) The global wavelet power spectrum (black line). The dashed line is the significance for the global wavelet spectrum, assuming the same significance level and background spectrum as in (b).